# Comment on "Rotary Photon Drag Enhanced by a Slow-Light Medium [1]"

ADRIAN C SELDEN[a*]

*The observations reported by Franke-Arnold et al (SCIENCE Reports, 1 July 2011 p. 65) do not provide evidence of slow light enhanced rotary photon drag as claimed, but arise from well-known saturable absorption phenomena consistent with the use of a spectrally broad light source, as described in their theoretical analysis (supporting online material)*

With reference to the article 'Rotary Photon Drag Enhanced by a Slow-Light Medium' [1], previous claims for the observation of 'slow light' via coherent population oscillations (CPO) in saturable absorption media [2, 3] have been invalidated both experimentally [4] and theoretically [5, 6] over the past several years. The reported phenomena (phase shift, pulse delay, modulation gain) merely reflect the finite relaxation time associated with saturable absorption i.e. 'slow response' rather than 'slow light', and can be interpreted by a standard theoretical description involving intensity driven absorption modulation [6, 7], as presented in the supporting online material [1]. The distinction can be traced to the original perturbation analysis, which predicts both broadband saturable absorption and narrow coherent hole burning when two light waves beat in a non-linear absorber, the coherent hole appearing when the frequency difference (beat frequency) is comparable with the inverse relaxation time viz. $\Delta\omega \sim 1/\tau$ [8]. Because both phenomena are tied to the relaxation time, which determines the response of the absorption *and* the frequency width of the coherent hole, the cause of the confusion is clear. Saturable absorption theory provides a good fit to the experimental data, as shown [1], but does not provide a theoretical basis for the 'slow light' interpretation. When the conditions for creating a narrow coherent hole in the absorption band are not met, there can be no reduction of group velocity and therefore no 'slow light' [5]. Conversely, both coherent hole burning (hence 'slow light') and saturable absorption can be observed simultaneously when *independent* narrow band pump and probe beams are combined to generate CPO in a saturable absorber [9].

More seriously, no independent test of image rotation – other than the apparent rotation of the elliptical beam of transmitted laser light – on which the claimed observation of 'rotary photon drag' is based, seems to have been made. Saturable absorption analysis suggests that the observed displacement merely reflects the angular displacement of peak transmission of the rotating ruby rod arising from the delayed response of the non-linear absorption. Image rotation could easily be tested by placing a wire grid in the incident beam and checking for rotation of the resulting diffraction pattern in the image plane. Secondly, the photon drag effect should rotate the plane of polarisation of the illuminating beam by an equal amount [10], which could be tested by employing a plane polarised laser beam in the current experiment. These simple tests are crucial to interpreting the observations, since saturable absorption theory suggests that *neither* of these rotations will be observed with the current experimental setup.

Saturable absorption is an inherently non-linear effect, whereas 'slow light' is a linear phenomenon, corresponding to a group velocity reduction inversely proportional to the width of the coherent hole [6]. The problem of distinguishing the two experimentally is challenging, particularly for relaxation times $\geq$ 10 ms, when the narrow frequency width of the coherent hole ($\leq$ 100 Hz) renders it difficult to detect. One way of avoiding this is to employ independent pump and probe beams at separate wavelengths within the broad absorption band [4]. In this way, it has been shown that modulating the pump intensity at one wavelength modulates the entire absorption band, imposing either a phase delay or phase advance on the intensity modulated probe, thereby simulating either 'slow' or 'fast' light [4] in circumstances where hole-burning allows only 'slow light' [3].

In light of the above considerations, one must therefore conclude that the observations described in [1] do not provide evidence of slow light enhanced rotary photon drag as claimed, but are well-known saturable absorption phenomena consistent with the use of a spectrally broad light source and a near-ideal saturable absorber (ruby) [1], as described in the theoretical analysis (supporting online material for [1]). It is recommended that the experiment be modified to allow a simple test of image rotation by incorporating a diffraction grid in the illuminating beam. Secondly, that a plane polarised incident beam be employed to test for rotation of the plane of polarisation, which should rotate through the same angle as the image [10]. Finally, it is suggested that consideration be given to designing a separate experiment where the 'slow light' propagation time ($\tau \leq 90$ μs in 6 mm length ruby rod [1]) can be observed unambiguously, rather than inferred indirectly (as here), to support the claims made for slow light enhancement of rotary photon drag.

[a]*former Visiting Professor*
*Department of Physics,*
*University of Zimbabwe*

*adrian_selden@yahoo.com